\documentclass{iaus}
\usepackage{graphicx}

\title[Discovery of the most rapidly rotating, magnetic massive star] 
{Discovery of the most rapidly-rotating, non-degenerate, magnetic massive star by the MiMeS collaboration}

\author[J.H. Grunhut et al.]   
{J.H. Grunhut$^{1}$, G.A. Wade$^1$, T. Rivinius$^2$, W.L.F. Marcolino$^3$, R.H.D. Townsend$^4$, and the MiMeS Collaboration}

\affiliation{$^1$Kingston, Canada; $^2$ESO Chile; $^3$Marseille, France; $^4$Madison, USA\\}

\pubyear{2010}
\volume{272}  
\pagerange{119--126}
\jname{Active OB stars}
\editors{A.C. Editor, B.D. Editor \& C.E. Editor, eds.}
\begin{document}

\maketitle

\begin{abstract}
We discuss the recent detection of a strong, organized magnetic field in the bright, broad-line B2V star, HD~142184, using the ESPaDOnS spectropolarimeter on the CFHT as part of the Magnetism in Massive Stars (MiMeS) survey. We find a rotational period of 0.50833 days, making it the fastest-rotating, non-degenerate magnetic star ever detected. Like the previous rapid-rotation record holder HR 7355 (also discovered by MiMeS: Oksala et al. 2010, Rivinius et al. 2010), this star shows emission line variability that is diagnostic of a structured magnetosphere.
\end{abstract}

\firstsection 
\section{Introduction \& observations}
Magnetic fields are unexpected in hot, massive stars due to the lack of convection in their outer envelopes. However, a small number of massive B stars host strong, organized magnetic fields, such as the chemically peculiar He strong stars (see Bohlender \& Monin, these proceedings) like the archetypical star $\sigma$~Ori~E (see Oksala et al., these proceedings) and the recently discovered B2V star HR~7355 (Oksala et al. 2010; Rivinius et al. 2010). These stars are rapidly rotating and host strong magnetic fields that are coupled to a co-rotating magnetosphere (Townsend, Owocki, Groote 2005).

Twenty-one high-resolution ($R\sim68000$) observations of the variable B2V star HD\,142184 were obtained with the ESPaDOnS spectropolarimeter at the Canada-France-Hawaii telescope between February and March 2010. These initial observations clearly show the presence of Zeeman signatures in the circular polarization, Stokes $V$ Least-Squares Deconvolved (LSD), mean line profiles, indicative of a magnetic field.

We also obtained six low-resolution ESO-VLT FORS observations in addition to 27 UVES observations in April 2010 to follow up the ESPaDOnS observations. 

\section{Rotational period, variability, and magnetic field geometry}
\begin{figure}
\centering
\includegraphics[width=5.2in]{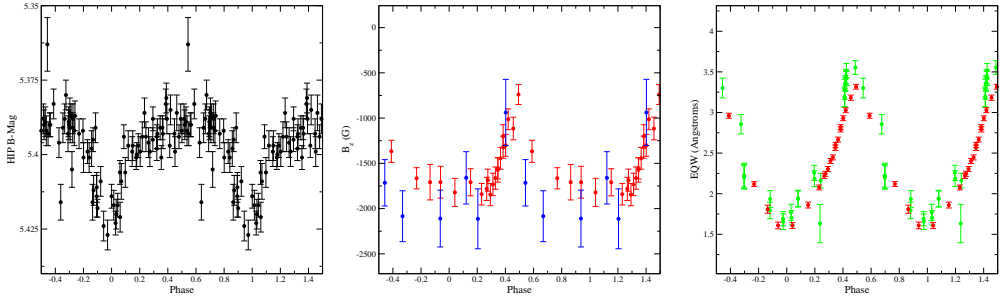}
\caption{Phased Hipparcos (left), longitudinal magnetic field (middle), and H$\alpha$ equivalent width measurements for HD 142184. Different colours correspond to the different instruments (red=ESPaDOnS, blue=FORS, green=UVES).}
\label{sidebyside}
\end{figure}

Using Hipparcos photometry, we find a single-wave period of $0.50831\pm0.00002$\,d (see Fig.~\ref{sidebyside}), which differs from the double-wave photometric light curve of HR 7355 or $\sigma$~Ori E, likely indicating a different geometry of the magnetic field. From our longitudinal magnetic fields measurements, we confirm this period, which we take to be the rotational period of this star (see Fig.~\ref{sidebyside}), making it the fastest rotating, non-degenerate, magnetic massive star! However, the period is sufficiently imprecise that the relative phasing between our current data and the Hipparcos data can be offset by so much as 0.5 cycles. Therefore, we have adopted a period of 0.50833\,d so that the relative phasing between the photometric minimum and the peak of the longitudinal field curve differs by 0.5 cycles - consistent with the predictions of semi-analytical models for a rotationally supported magnetosphere (Townsend 2008; Townsend \& Owocki 2005).

From NLTE model fits and a Fourier analysis to the ESPaDOnS spectra, we find that $T_{\rm eff}=19\pm2$\,kK, $\log(g)=2.95\pm0.04$, and $v\sin i=270\pm10$\,km\,s$^{-1}$.

\begin{figure}
\centering
\includegraphics[width=1.6in]{s2-09_grunhut_fig2.eps}
\includegraphics[width=1.6in]{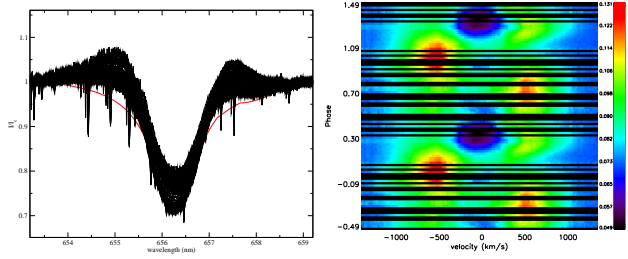}\\
\caption{{\bf Left:} H$\alpha$ (black) profiles for different nights compared to a NLTE model profile (red). {\bf Right:} Phased residual H$\alpha$ variations relative to the NLTE model.}
\label{ha_dyn}
\end{figure}

In addition to the photometric and magnetic periodicity, we also find that H$\alpha$ varies with the same period, as shown in Fig.~\ref{sidebyside}. H$\alpha$ shows line profile variations of emission extending to high velocities, as shown in Fig.~\ref{ha_dyn}. The double-lobed pattern and equivalent width variations strongly suggests that HD 142184 hosts a structured magnetosphere similar to $\sigma$~Ori~E and HR 7355 consisting of co-rotating, magnetically confined clouds of stellar wind plasma.

Assuming rigid rotation, we infer that the inclination $i\sim75^{\circ}$. From fits to the longitudinal field curve shown in Fig.~\ref{sidebyside} we estimate that HD 142184 hosts a mainly dipole magnetic field, with a strength at its pole of $\sim20$\,kG, and a magnetic axis nearly aligned with the rotation axis. Currently, the inclination is poorly constrained, which results in a large possible range for the dipole field strength. However, using the predictions of the Rigidly Rotating Magnetosphere model (Townsend \& Owocki 2005), we expect to better constrain the geometry of HD 142184 based on the variations shown in Fig.~\ref{sidebyside}.


\begin{thebibliography}{}
\bibitem[Oksala et al.(2010)]{oksala10}Oksala et al., 2010, \textit{MNRAS}, 405, L510
\bibitem[Rivinius et al.(2010)]{rivi10}Rivinius et al., 2010, \textit{MNRAS}, 405, L46
\bibitem[Townsend \& Owocki(2005)]{town05}Townsend, R.H.D., Owocki, S.P., 2005, \textit{ApJ}, 630, L81
\bibitem[Townsend, Owocki, Groote(2005)]{}Townsend, R.H.D., Owocki, S.P., Groote, D., 2005, ApJ, 630, L81
\bibitem[Townsend(2008)]{town08}Townsend, R.H.D., 2008, \textit{MNRAS}, 389, 559
\end{thebibliography}
\end{document}